\documentclass{article}
\usepackage{spconf}
\ninept
\usepackage{mathtools, cuted}
\usepackage{setspace}
\usepackage{setspace}
\usepackage{enumitem}
\usepackage[utf8]{inputenc}
\usepackage{amsmath,amsfonts,amssymb}
\usepackage{amsthm}
\usepackage{bm}%
\usepackage{graphicx,graphics}
\usepackage{graphicx}
\usepackage{verbatim}
\usepackage{cases}%
\usepackage[noadjust]{cite}%
\usepackage{color,xcolor}
\usepackage{cite,url}
\usepackage{algorithm}
\usepackage[noend]{algpseudocode}
\usepackage{balance}
\usepackage{multirow}
\usepackage{stfloats}
\usepackage{xspace}
\usepackage{siunitx}
\usepackage{subfig} 
\usepackage{mathtools} 
\usepackage{url}
\usepackage{float}
\usepackage{caption}
\usepackage{tikz}
\usepackage{subfig}

\begin{document}
\title{Identification and Quantification of Aerosol Hot-spots over Lahore Region using MODIS Data}

	\name{Safa~Ashraf \quad\quad Zubair Khalid \quad\quad Muhammad Tahir \quad\quad Momin Uppal}

\address{Department of Electrical Engineering, School of Science and Engineering\\
Lahore University of Management Sciences, Lahore 54792, Pakistan}

\maketitle

\begin{abstract}
The increased concentration of aerosols in the air caused by ever-rising urbanization and the development of various industries has horrendous consequences on human health, environment and climate. The first step to counter adverse effects of air pollution in any region is to identify locations with a high concentration of aerosols, termed aerosol hot-spots. We specifically focus on the region of Lahore, Pakistan, a city that has consistently been ranked among the top ten most polluted cities in the world. In order to identify aerosol hot-spots in the city, we utilize two-year (2017-18) Aerosol Optical Thickness (AOT) data from MODerate resolution Imaging Spectroradiometer (MODIS) on the Aqua satellite. We propose a method based on the Glowworm Swarm Optimization (GSO) algorithm that discovers several aerosol hot-spots over Lahore comprising of major highways and industrial areas. Furthermore, we formulate two quantification metrics to gauge the amount of aerosol content over each hot-spot. We also analyze the temporal variation of aerosol content that suggests the addition or suppression of pollution sources in the hot-spots. Unlike previous studies, our work provides novel insight into the regional aerosol concentration of Lahore and calls for maintenance of air quality of the regions falling under the identified aerosol hot-spots.
\end{abstract}


%

\section{Introduction}
Many natural phenomena as well as anthropogenic activities are responsible for creating an unhealthy change in the composition of air. However, degraded air quality in urbanized areas of developing countries such as Pakistan is primarily caused by anthropogenic sources. Air pollutants cause dreadful environmental conditions like contamination of crops and livestock, production of greenhouse gases, acid rain, reduced visibility and formation of smog \cite{1}. Some pollutants can even alter cloud properties by absorbing and reflecting sunlight, due to which climate and the hydrological cycle is affected \cite{2}. Its impact of highest concern however is on the human body with cardiopulmonary health being the biggest sufferer \cite{70}. The severity of the air pollution crisis in Pakistan is reflected in the World Air Quality Report compiled by IQAir AirVisual and Greenpeace Southeast Asia in 2018, according to which, two major cities of Pakistan, Lahore and Faisalabad, have been listed among the top ten polluted cities of the world\footnote{\url{https://www.airvisual.com/.}}. Moreover, according to the 2015 World Health Organization (WHO) report, almost 60,000 people in the country died from large amounts of fine particles in the air. This is considered one of the highest death tolls recorded globally due to air pollution~(FAO of UN report, 2018). On top of these, according to a 2014 report by World Bank~\cite{3}, more than 20,000 premature deaths among adults and almost 5,000,000 cases of illness among children are recorded in Pakistan each year due to degraded air quality.

The aforementioned hazards and figures are more than enough of a reason to motivate research and analyses followed by policy decisions to counter harmful effects of air pollution. Researchers have been trying to study air pollution trends in Pakistan over the last two decades using both satellite and ground-based sensors. In a work published recently, seasonal trend in the properties of aerosols based on Aerosol Optical Thickness (AOT) data from MODerate resolution Imaging Spectroradiometer (MODIS) and ground-based AErosol RObotic NETwork (AERONET) station was studied in Lahore along with the classification of aerosols~\cite{6}. Furthermore, MODIS has been used to identify hotspots in different cities~(e.g.,\cite{kumari2020modis,lizundia2020spatio}). Spatio-temporal variations in aerosol concentration for several cities of Pakistan have been analyzed using the Hybrid Single Particle Lagrangian Integrated Trajectory (HYSPLIT) model on satellite-based data of MODIS, Multi-angle Imaging Spectroradiometer (MISR) and Total Ozone Mapping Spectrometer (TOMS) via backtracking air mass trajectories~\cite{36,37,38}. Trends of various aerosol parameters have been studied on a country-wide scale using data from several satellites and AERONET~\cite{69}. Heavy pollution episodes were classified into dust episode~(DE) and haze episode~(HE) over Karachi and Lahore using the correlation between AOT and Angstrom Exponent (AE) of MODIS and AERONET data~\cite{34}. Based solely on the ground-based AERONET data, an elaborate study of various aerosol properties was carried out over Lahore \cite{31}. In contrast to AERONET, source apportionment and chemical characterization of aerosols have been performed using Particulate Matter (PM) samplers and air quality monitors in Lahore~\cite{64,63,68} and in a few other urban cities of Pakistan~\cite{65}. Source sector contributions in the country have been studied using simulation based on Weather Research and Forecasting Model coupled with Chemistry (WRF-Chem)~\cite{66}. Although these studies provide useful results and improve the understanding of air pollution crisis in Pakistan, they miss out on identifying air pollution sources in terms of their locations and aerosol content in a particular city.

In this paper, we present a spatial characterization of AOT for the city of Lahore in Pakistan. To develop this characterization, we identify aerosol hot-spots using a method based on Glowworm Swarm Optimization (GSO)~\cite{24} algorithm on a two-year~(2017-2018) AOT data from MODIS mounted on Aqua, an Earth Science satellite mission of NASA.  It is
noted that a modified version of GSO algorithm has been proposed to model the back-tracking of aerosols in Chengdu city of China for the purpose of identifying air pollution sources on the scale of factories~\cite{15}. In contrast to model back-tracking, the use of GSO in this study is for the optimization of AOT data for aerosol hot-spot identification in Lahore. Aerosol hot-spots in this work are defined as those regions where high AOT values are observed relative to their surroundings. It is important to identify locations of these hot-spots since they not only suffer from high aerosol concentration but also act as sources of air pollution and disperse the aerosols in their vicinity. Our proposed method discovers eleven aerosol hot-spots around the city, mostly comprising industrial areas and major urbanized highways. Furthermore, we formulate two quantification metrics that gauge aerosol content over these identified hot-spots. The temporal trend of aerosol content is also carried out to infer the addition or suppression of pollution sources within each hot-spot.

\section{Preliminaries}
\subsection{Study Site}
Lahore is the capital city of the province of Punjab, situated in the northeast region of Pakistan. According to the UN World Urbanization Prospects, the latest estimate of population of the city and adjacent suburban areas in the year 2019 is that of 12,188,196 and is considered the second most populous city of the country \footnote{\url{https://population.un.org/wup/.}}. Vehicular manufacturing, steel, chemicals, construction materials, pharmaceuticals, flour and rice mills make up its major industries. Increased aerosol concentration in Lahore is a consequence of secondary Particulate Matter (PM), diesel emissions, biomass burning, coal combustion, two-stroke vehicle exhaust, industrial sources and re-suspended dust~\cite{62,63}. Figure~\ref{ident}(a) shows the area of Lahore for which all the analyses are carried out. It consists of the main city of Lahore, small towns in the outskirts, River Ravi, some field areas and several highways that connect Lahore to other cities.
	

\subsection{Data Description}
Since the ground-based sensors are sparsely present in Lahore, satellite-based data is an attractive alternative to carry out research due to its global coverage. In this work, daily AOT data from the MODIS of Aqua satellite available in the spatial resolution of \SI{3}{\kilo\metre} and temporal resolution of 1-2 days, named as MYD04\_3K, associated with the years 2017 and 2018 is used. This product uses Dark Target algorithm \cite{11} to retrieve AOT and is shown to be positively correlated with the AOT measurements of the ground-based AERONET in Lahore \cite{28,9,30}. The data is downloaded from the online LAADS DAAC data archive \footnote{\url{https://ladsweb.modaps.eosdis.nasa.gov/search/.}}.

\let\vec\mathbf
\section{Identification of Aerosol Hot-spots}
The problem of identifying the locations of aerosol hot-spots in a region is equivalent to the problem of locating all the local maxima of the AOT data that is two-dimensional, multi-modal and discrete in nature. The solution of this problem requires application of a multi-modal optimization method. Multi-modal functions usually model signals with multiple sources like sound, heat, light and leaks in pressurized systems and are optimized well with Swarm Intelligence~(SI) techniques \cite{26,25}.

\subsection{Glowworm Swarm Optimization Algortihm}
GSO algorithm is an SI technique that enables a population of agents called glowworms to converge at the local maxima through interaction with each other as well their environment. These glowworms possess a luciferin-level (analogous to the glow in natural glowworms) which plays a vital role in the algorithm's progression. GSO is different from the earlier approaches to multi-modal optimization due to its adaptive decision range~(to be discussed shortly). Moreover, it is memoryless, gradient free and does not require knowledge of the global information of the function to be optimized~\cite{25}. GSO algorithm is governed by the following three mechanisms~\cite{24}:

\begin{itemize}[noitemsep]
	\item \emph{Fitness Broadcast}: Luciferin-level, denoted by $l_{i}$ for every $i^{\rm th}$ glowworm, takes its value based on the fitness of its location in the search space. Intuitively, the closer a glowworm is to a local maxima, the higher the value of its luciferin-level. This value is broadcast to all glowworms in the search space.
	\item \emph{Adaptive Neighborhood}: For each glowworm, there is an adaptive local decision range which acts like its vision and is used to form its neighborhood consisting of other glowworms. It is based on a variable range $r_{d}^{i}$ that is bounded by a hard-limited sensor range $r_{s}$, such that $0<r_{d}^{i}<r_{s}$. The range $r_{d}^{i}$ is updated at every iteration. This adaptive nature of neighborhood tries to ensure that all local peaks in the data are sought.
	\item \emph{Positive Taxis}: Among all the neighbors of a glowworm, the glowworm moves towards a neighbor that is chosen based on a probabilistic heuristic. Since this movement is towards the stimulus~(higher luciferin-level), the mechanism is termed as the positive taxis.
\end{itemize}

Details of the method proposed to identify aerosol hot-spots using GSO algorithm is described next. Note that the following method is applied to the AOT data of each day separately.

\begin{figure*}[!ht]
	\centering
	\subfloat[Study site~(Courtesy: Google Maps).]
	{
		\includegraphics[scale=0.43]{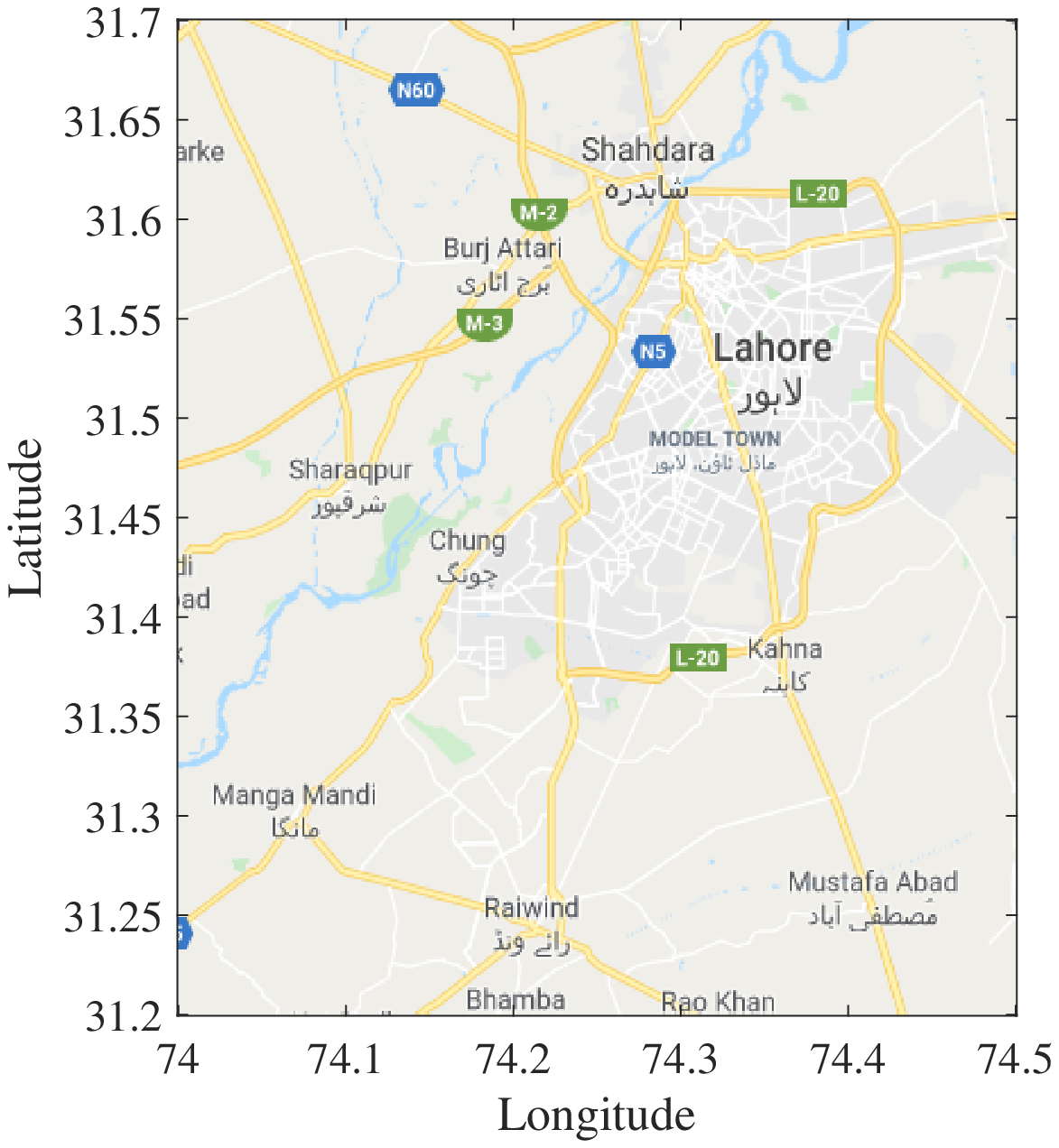}
	}
    \hspace{2mm}
	\subfloat[Identified aerosol hot-spots (Year 2017).]
	{
		\includegraphics[scale=0.43]{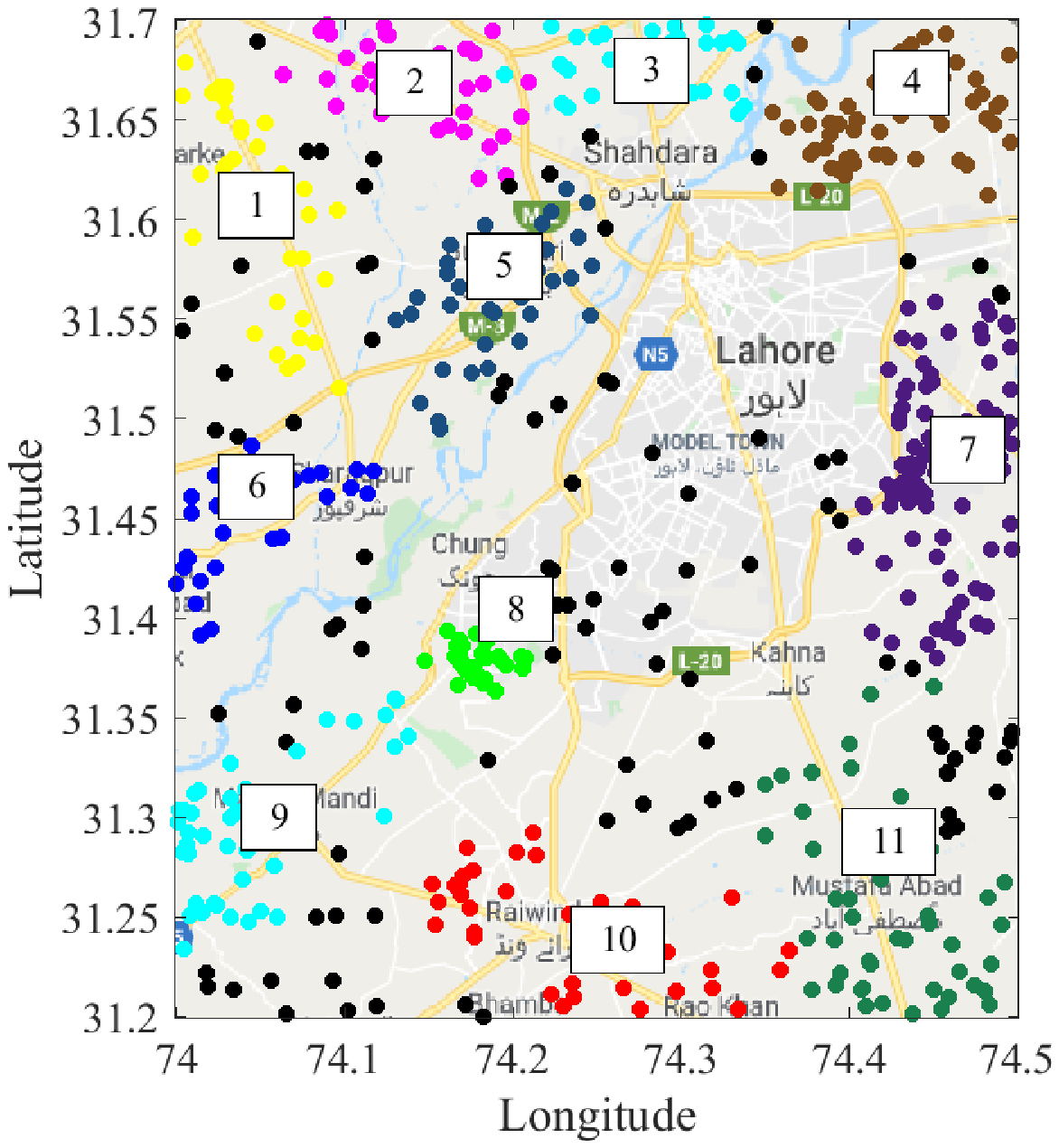}
	}
	\hspace{2mm}
	\subfloat[Identified aerosol hot-spots (Year 2018).]
	{
		\includegraphics[scale=0.43]{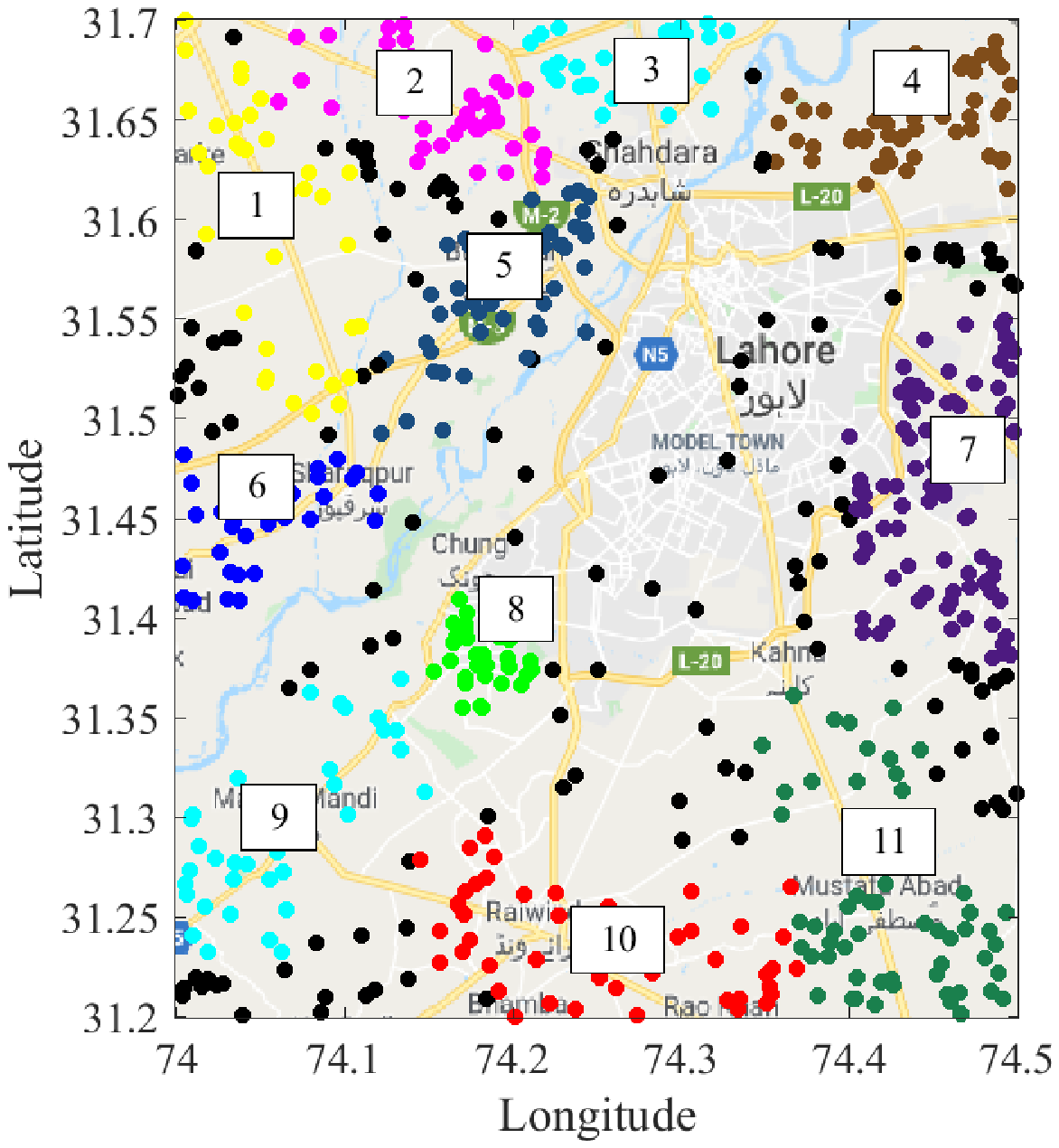}
	}
	\caption{Identification of Aerosol Hot-spots using GSO over Lahore}
	\label{ident}
\end{figure*}

\subsubsection{Initialization}
Each data point in AOT data of a day is considered a glowworm and a default luciferin-level $l_{0}$ is assigned to it. Other parameters, to be described shortly, that govern the GSO algorithm are also initialized in this step. The next three steps of the algorithm are repeated for 200 iterations that are sufficient to ensure that the glowworms converge to the local maxima of each day.

\subsubsection{Iterative Steps of Algorithm}
Luciferin-level of each glowworm represents the closeness of that glowworm to a local maxima. Since the objective function is the two-dimensional MODIS AOT data for the day under consideration, the glowworms at the coordinates with high AOT value get a higher luciferin-level based on the principle
\begin{equation}
\label{luci}
l_{i}(t+1) = (1 - \rho) l_{i}(t) + \gamma y_{i},
\end{equation}
where $l_{i}(t)$ is the luciferin-level of an $i^{\rm th}$ glowworm at iteration $t$, $y_{i}$ is the $i^{\rm th}$ AOT value from the data of a day, $\rho$ is the luciferin decay constant which controls how much of the previous luciferin-level has to be retained, $\gamma$ is luciferin enhancement constant that scales the luciferin-level with the function value and $t$ is the iteration number.

Next comes the movement phase that guides the AOT points to move towards the local solutions. This is achieved by the AOT points moving towards a neighbor that has a higher luciferin-level based on a probabilistic heuristic. The heuristic $p_{ij}(t)$ represents the probability with which an $i^{\rm th}$ AOT point will move towards a $j^{\rm th}$ AOT point and is defined as
\begin{equation}
p_{ij}(t) = \frac{l_{j}(t)-l_{i}(t)}{ \sum\limits_{k\in N_{i}(t)} l_{k}(t)-l_{i}(t)},
\end{equation}
where $j$ belongs to $N_{i}(t)$ with $N_{i}(t)=\{j:d_{ij}(t) < r_{d}^{i}; l_{i}(t)<l_{j}(t)\}$ is a set of neighbors of an AOT point $i$ that are closer to the solution than the $i^{\rm th}$ point and are located at a Euclidean distance of less than $r_{d}^{i}$, the adaptive decision range for the $i^{\rm th}$ point. Here, $d_{ij}(t)$ denotes the Euclidean distance between $i^{\rm th}$ and $j^{\rm th}$ AOT points at the iteration $t$. At iteration number $t$, the update equation for the coordinates of each AOT point is given by
\begin{equation}
x_{i}(t+1) = x_{i}(t) + s \frac{x_{j}(t)-x_{i}(t)}{\|x_{j}(t)-x_{i}(t)\|},
\end{equation}
where $x_{i}(t) \in \mathbb{R}^{2}$ represents the coordinates (latitude and longitude) of $i^{\rm th}$ AOT data point at iteration $t$. In this step, the $i^{\rm th}$ AOT point chooses a glowworm $j$ from $N_{i}(t)$ randomly with a probability $p_{ij}(t)$. $\|\cdot\|$ is the Euclidean norm operator and $s$ is a positive quantity that denotes the step size.

Before starting the next iteration from \eqref{luci} again, each AOT point needs to update the value of decision range $r_{d}^{i}$ for forming a neighborhood in the next iteration. The decision range of an $i^{\rm th}$ AOT point is updated based on the following rule
\begin{equation}
r_{d}^{i}(t+1) = \min \{r_{s}, \max \{0, r_{d}^{i}(t) + \beta(n_{t} - |N_{i}(t)|)\}\},
\end{equation}
where $r_{d}^{i}(0)$ is initialized at the first iteration as $r_{0}=0.2$ for all $i$. $r_{s}$ is the hard-limited sensor range. The parameter $\beta$ affects the rate of change of the neighborhood range. $n_{t}$ is the neighborhood threshold that indirectly controls the number of neighbors by influencing the neighborhood range at every iteration. $|N_{i}(t)|$ measures total number of neighbors of the $i^{\rm th}$ AOT point in the current iteration.

\begin{figure*}[!ht]
	\centering
	\subfloat[Regional AOT around aerosol peaks (2017).]
	{
		\includegraphics[scale=0.41]{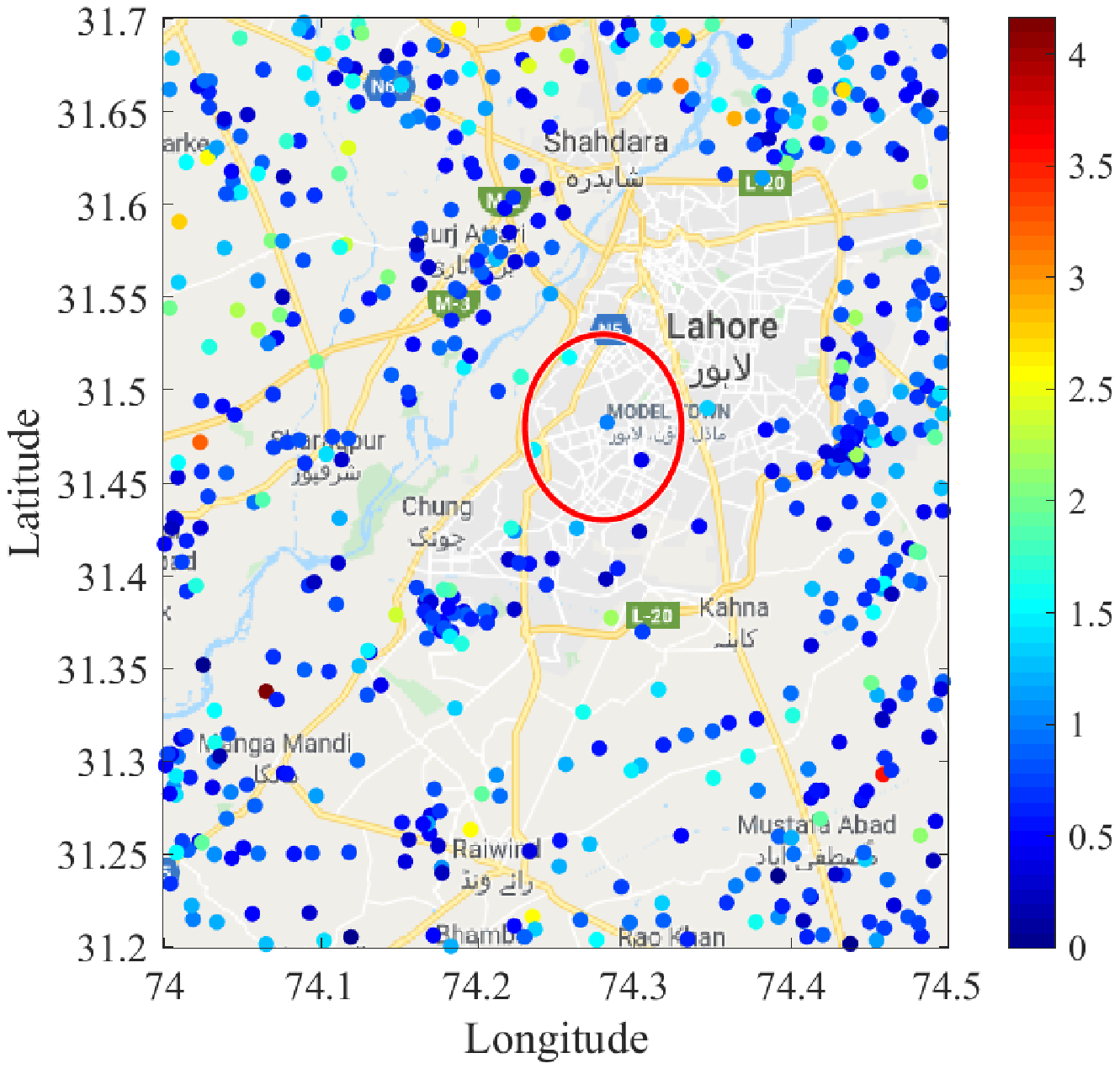}
	}
	\hspace{2mm}
	\subfloat[Regional AOT around aerosol peaks (2018).]
	{
		\includegraphics[scale=0.41]{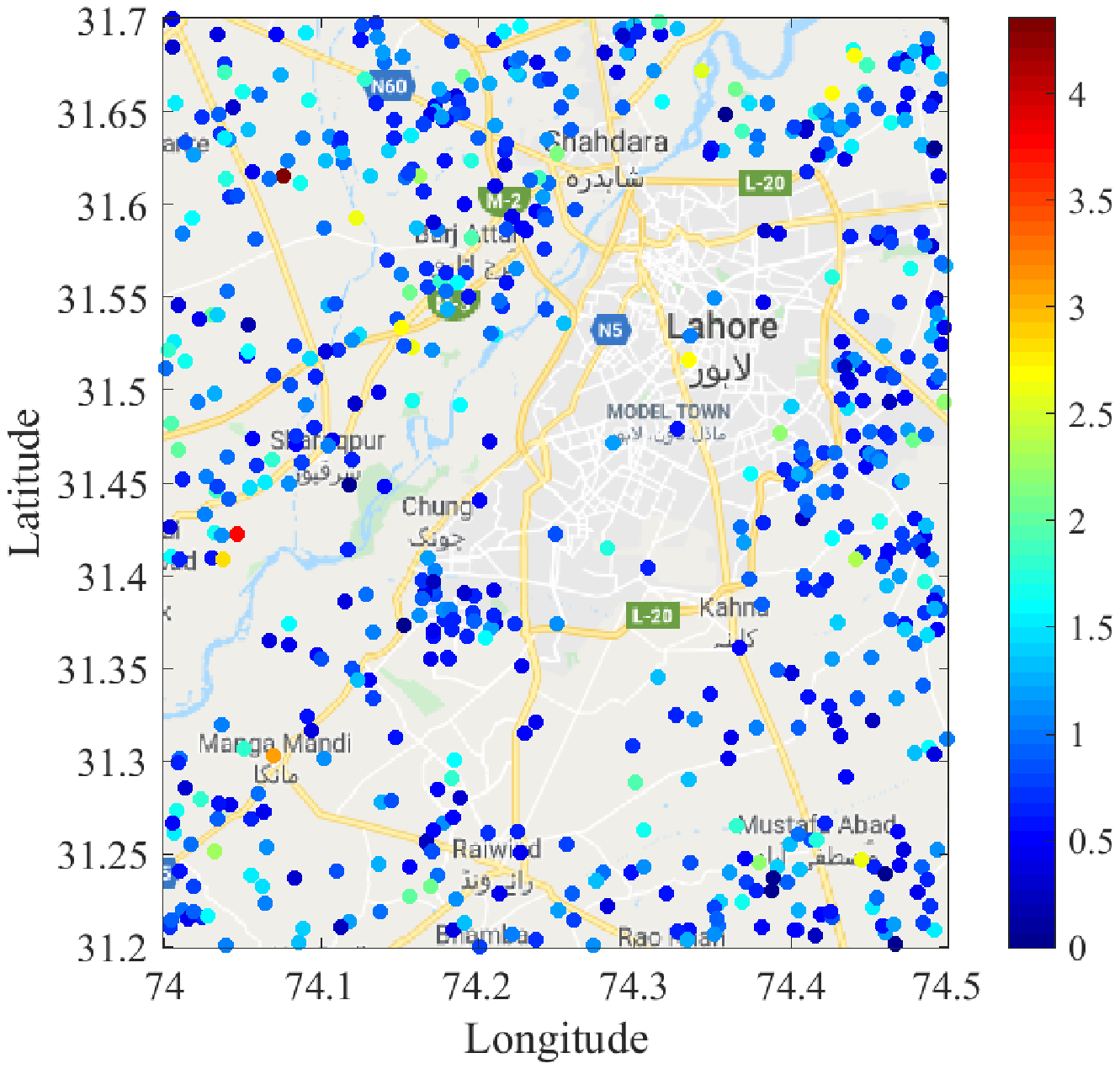}
	}
	\hspace{2mm}
	\subfloat[Average AOT over hot-spots.]
	{
		\includegraphics[scale=0.38]{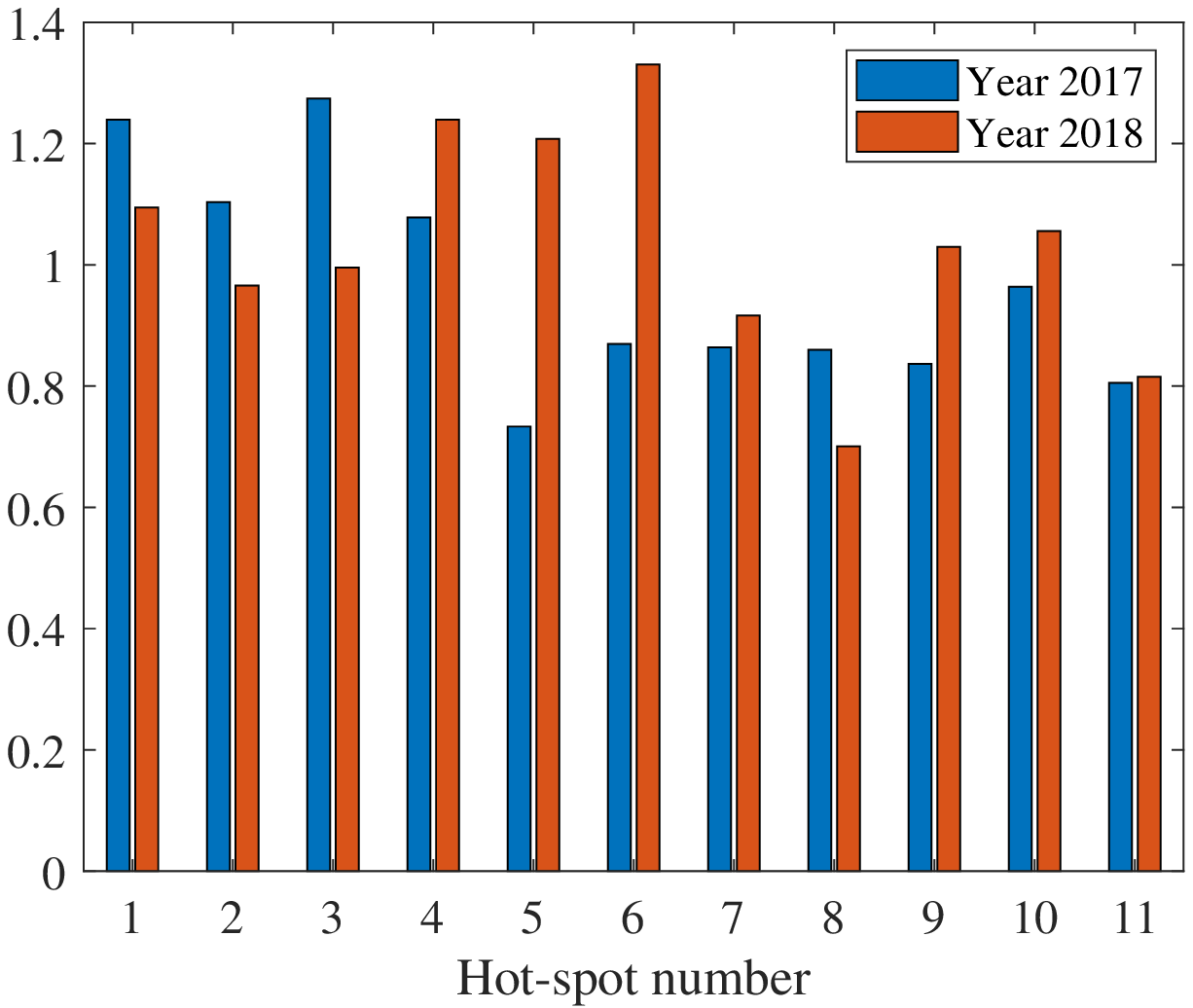}
	}
	\caption{Quantification of Aerosol Content over Identified Hot-spots}
	\label{quant}
\end{figure*}

\subsection{Identification of Aerosol Hot-Spots}
To identify the aerosol hot-spots in the region of study, we propose to initialize the parameters of the GSO algorithm with the recommended values given in \cite{25}. Similar to $\rho$ and $l_{0}$, the sensor range $r_{s}=0.2$ is chosen based on multiple experimental trials. The initialized parameters used in our experiments is summarized in Table \ref{tab:params-aot}.

\begin{table}[h]

		\begin{center}
			\begin{tabular}{||c c c c c c||}
				\hline
				$\rho$ & $\gamma$ & $\beta$ & $n_{t}$ & $s$ & $l_{0}$  \\ [0.5ex]
				\hline\hline
				0.2 & 0.6 & 0.08 & 5 & 0.03 & 2 \\[1ex]
				\hline
			\end{tabular}
		\caption{Initialized values of algorithm's parameters}
		\label{tab:params-aot}
		\end{center}
\end{table}

\noindent Using the above initialization of the GSO algorithm on the data for each day, AOT data points that are spread across the whole region form clusters. The centroid of these clusters happen to be the local maxima of the daily AOT data being optimized. Therefore, we determine the local maxima or the locations of the daily aerosol peaks by calculating the coordinates of these cluster centroids. To obtain the desired aerosol hot-spots of the region, we propose to repeat this method of localization for all days of the two years separately and overlay the coordinates of the identified aerosol peak locations. Based on the information from geographical map of the region under study, shown in the Figure~\ref{ident}(a), we map these hot-spots to geographical locations manually and color-code them for convenient visualisation.

The aerosol hot-spots found by employing the proposed method are illustrated in Figure~\ref{ident}(b) and \ref{ident}(c) in different colors for the two years. Note that the peak locations shown in black are not mapped to any aerosol hot-spot. The regions that fall within the 11 labeled hot-spots indicated on the figures are listed below.

\begin{enumerate}[noitemsep]
	\item Sheikhupura-Sharaqpur Road
	\item Lahore-Sheikhupura Road
	\item Ferozewala and Surroundings
	\item Shahdara Reserve Forest, Lakhodair Landfill and surrounding Field Areas
	\item Lahore-Jaranwala Road~(Shahdara to Sharaqpur)
	\item Lahore-Jaranwala Road~(Mandi Faizabad to Sharaqpur)
	\item Areas around Barki Road and DHA Phases 5, 6 and 8, Residential/Field Areas around Ruhi Nala Road
	\item Industrial Area near Valencia and Bahria Town
	\item Multan Road (Manga Mandi to Phool Nagar)	
	\item Around Sundar Industrial Estate and Kasur-Raiwind Road
	\item Around Lahore-Kasur Road
	
\end{enumerate}

The regions identified in the hot-spots comprise of industrial areas, major urbanized highways connecting Lahore to nearby cities, and open field areas. This finding suggests plausibility of the employed method for the identification of aerosol hot-spots in a region. The recognition of aforementioned aerosal hot-spots calls for appropriate preventive measures to set right and maintain poor air quality of the discerned areas.

\section{Quantification of Aerosol Content}
In this section, we formulate two basic quantification metrics based on spatial and temporal averaging to quantify and visualize the aerosol content in the region of study.

\subsection{Regional AOT around Aerosol Peaks}
After we determine the locations of the aerosol peaks from the daily AOT data over the period of two years, we compute a value of regional AOT in the vicinity of each identified peak. This vicinity is characterized by a radius of 0.05 decimal degrees on the geographic coordinate system, which covers an area of \SI{9}{\kilo\metre}${}^2$, is chosen to incorporate the surrounding AOT points of an aerosol peak for spatial averaging. We define the regional AOT around the $i^{\rm th}$ aerosol peak as
\begin{equation}
{\rm AOT}_{i} = \frac{1}{M_{i}} \sum_{j\in R_{i}} y_{ij},
\end{equation}
where $R_{i} = \{j: d_{ij} \leq 0.05\}$, $d_{ij}$ is the Euclidean distance between the locations of $i^{\rm th}$ aerosol peak and $j^{\rm th}$ AOT point of the daily AOT data in which the peak is sought, $M_{i}$ is the number of AOT points in the region $R_{i}$ and $y_{ij}$ is the AOT value of the $j^{\rm th}$ AOT point in $R_{i}$. The regional AOT of each aerosol peak is shown in Figure \ref{quant}(a) and \ref{quant}(b). Note that the red circle in Figure \ref{quant}(a) is shown to represent the radius for spatial averaging corresponding to the aerosol peak at its center. From these figures, one can easily point out the peak locations with high concentration of aerosol. The areas around these locations correspond to poor air quality in terms of aerosols.

\subsection{Average AOT over Aerosol Hot-spots}
To quantify the average aerosol content within a hotspot, in contrast to around an aerosol peak, average AOT over an $i^{\rm th}$ hot-spot, denoted by $\overline{\rm AOT_{i}}$, is defined as
\begin{equation}
\overline{\rm AOT_{i}} = \frac{1}{K_{i}}\sum_{j\in H_{i}} {\rm AOT_{j}},
\end{equation}
where $K_{i}$ denotes the number of aerosol peaks in the $i^{\rm th}$ hot-spot $H_{i}$. This quantification metric measures the average aerosol content within a hot-spot $H_{i}$ by averaging the regional AOT of all aerosol peaks in $H_{i}$. Figure \ref{quant}(c) shows the average AOT over the identified hot-spots for the two years. From this quantification, conclusions can be drawn about the severity of air pollution within each hot-spot. Moreover, temporal trend of aerosol concentration associated with each hot-spot can be inferred by comparing the values for the two years. For example, the aerosol concentration of hot-spot 5 and 6, both consisting of regions on the Lahore-Jaranwala Road show a significant amount of increase in the value with time. This may indicate new pollution sources that are not present in the year 2017 in these regions. However, as a future direction, we suggest to consider AOT data of several years at a time to make this specific analysis less simple.

\section{Conclusions}
The first step to control the crisis of poor air quality in an urbanized area like Lahore is to identify locations that posses high concentration of aerosols (termed as aerosol hot-spots). To achieve this objective, we have carried out identification of aerosol hot-spots in Lahore using a method based on GSO algorithm on the \SI{3}{\kilo\metre} AOT data of two years (2017-2018) from Aqua MODIS. Eleven aerosol hot-spots have been identified around the city of Lahore comprising of industrial areas and urbanized highways which suggests validity of our proposed method. To gauge the severity and assess the temporal variation of aerosol content over the hot-spots, we have also proposed two quantification metrics, regional AOT around aerosol peaks and average AOT over hot-spots. Our work provides a useful aerosol distribution of the city of Lahore and calls for timely air quality maintenance of the regions that lie within the identified aerosol hot-spots.

\bibliography{IEEEabrv,ref}

\end{document}